\documentclass[conference]{IEEEtran}

\usepackage{listings}
\usepackage{graphicx}
\usepackage{caption}
\usepackage{subcaption}

\usepackage[utf8]{inputenc} 
\usepackage[T1]{fontenc}    
\usepackage{hyperref}       
\usepackage{url}            
\usepackage{booktabs}       
\usepackage{amsfonts}       
\usepackage{nicefrac}       
\usepackage{microtype}      
\usepackage{xcolor}         
\usepackage{tcolorbox}
\usepackage{float}
\usepackage{amsmath}
\usepackage{tabularx}
\usepackage{multirow}

\definecolor{orange}{rgb}{1,0.5,0}
\definecolor{Gray}{gray}{0.9}
\definecolor{revision}{RGB}{0,0,255}
\definecolor{cellgrey}{RGB}{171,171,171}

\definecolor{primekeywords}{rgb}{0.13,0.13,1}
\definecolor{secondkeywords}{rgb}{0,0.5,0}
\definecolor{codebackground}{rgb}{1,1,1}
\definecolor{greycomments}{rgb}{0.6,0.6,0.6}
\definecolor{ruleColor}{rgb}{0.1, 0.3, 0.1}%

\newcommand{\ignore}[1]{}

\newcommand{\revisiondel}[1]{}

\newcommand{\revisionstart}{\begin{color}{revision}}
\newcommand{\revisionend}{~\!\!\end{color}}

\newcommand{\sysname}{SC-Bench}

\makeatletter
\newenvironment{btHighlight}[1][]
{\begingroup\tikzset{bt@Highlight@par/.style={#1}}\begin{lrbox}{\@tempboxa}}
{\end{lrbox}\bt@HL@box[bt@Highlight@par]{\@tempboxa}\endgroup}

\newcommand\btHL[1][]{%
  \begin{btHighlight}[#1]\bgroup\aftergroup\bt@HL@endenv%
}
\def\bt@HL@endenv{%
  \end{btHighlight}%
  \egroup
}
\newcommand{\bt@HL@box}[2][]{%
  \tikz[#1]{%
    \pgfpathrectangle{\pgfpoint{1pt}{0pt}}{\pgfpoint{\wd #2}{\ht #2}}%
    \pgfusepath{use as bounding box}%
    \node[anchor=base west, fill=yellow!75,outer sep=0pt,inner xsep=1pt, inner ysep=0pt, rounded corners=3pt, minimum height=\ht\strutbox+1pt,#1]{\raisebox{1pt}{\strut}\strut\usebox{#2}};
  }%
}
\newcommand{\eg}{{\textit{e.g.}}}
\newcommand{\ie}{{\textit{i.e.}}}

\newcommand{\boldparagraph}[1]{\vspace{1ex}\noindent{\textit{\textbf{#1}}}}
\newcommand{\italicparagraph}[1]{\underline{\textit{#1}}}

  {\begin{tcolorbox}[colback=gray!10]\small}%
  {\end{tcolorbox}}

\lstdefinelanguage{Rust}{%
  sensitive%
, morecomment=[l]{//}%
, morecomment=[s]{/*}{*/}%
, moredelim=[s][{\itshape\color[rgb]{0,0,0.75}}]{\#[}{]}%
, morestring=[b]{"}%
, alsodigit={}%
, alsoother={}%
, alsoletter={!}%
, morekeywords={break, continue, else, for, if, in, loop, match, return, while}  
, morekeywords={as, const, let, move, mut, ref, static}  
, morekeywords={dyn, enum, fn, impl, Self, self, struct, trait, type, union, use, where}  
, morekeywords={crate, extern, mod, pub, super}  
, morekeywords={unsafe}  
, morekeywords={abstract, alignof, become, box, do, final, macro, offsetof, override, priv, proc, pure, sizeof, typeof, unsized, virtual, yield}  
, morekeywords=[2]{Add, AddAssign, Any, AsciiExt, AsInner, AsInnerMut, AsMut, AsRawFd, AsRawHandle, AsRawSocket, AsRef, Binary, BitAnd, BitAndAssign, Bitor, BitOr, BitOrAssign, BitXor, BitXorAssign, Borrow, BorrowMut, Boxed, BoxPlace, BufRead, BuildHasher, CastInto, CharExt, Clone, CoerceUnsized, CommandExt, Copy, Debug, DecodableFloat, Default, Deref, DerefMut, DirBuilderExt, DirEntryExt, Display, Div, DivAssign, DoubleEndedIterator, DoubleEndedSearcher, Drop, EnvKey, Eq, Error, ExactSizeIterator, ExitStatusExt, Extend, FileExt, FileTypeExt, Float, Fn, FnBox, FnMut, FnOnce, Freeze, From, FromInner, FromIterator, FromRawFd, FromRawHandle, FromRawSocket, FromStr, FullOps, FusedIterator, Generator, Hash, Hasher, Index, IndexMut, InPlace, Int, Into, IntoCow, IntoInner, IntoIterator, IntoRawFd, IntoRawHandle, IntoRawSocket, IsMinusOne, IsZero, Iterator, JoinHandleExt, LargeInt, LowerExp, LowerHex, MetadataExt, Mul, MulAssign, Neg, Not, Octal, OpenOptionsExt, Ord, OsStrExt, OsStringExt, Packet, PartialEq, PartialOrd, Pattern, PermissionsExt, Place, Placer, Pointer, Product, Put, RangeArgument, RawFloat, Read, Rem, RemAssign, Seek, Shl, ShlAssign, Shr, ShrAssign, Sized, SliceConcatExt, SliceExt, SliceIndex, Stats, Step, StrExt, Sub, SubAssign, Sum, Sync, TDynBenchFn, Terminal, Termination, ToOwned, ToSocketAddrs, ToString, Try, TryFrom, TryInto, UnicodeStr, Unsize, UpperExp, UpperHex, WideInt, Write}
, morekeywords=[2]{Send}  
, morekeywords=[3]{bool, char, f32, f64, i8, i16, i32, i64, isize, str, u8, u16, u32, u64, unit, usize, i128, u128}  
, morekeywords=[4]{Err, false, None, Ok, Some, true}  
, morekeywords=[3]{AccessError, Adddf3, AddI128, AddoI128, AddoU128, ADDRESS, ADDRESS64, addrinfo, ADDRINFOA, AddrParseError, Addsf3, AddU128, advice, aiocb, Alignment, AllocErr, AnonPipe, Answer, Arc, Args, ArgsInnerDebug, ArgsOs, Argument, Arguments, ArgumentV1, Ashldi3, Ashlti3, Ashrdi3, Ashrti3, AssertParamIsClone, AssertParamIsCopy, AssertParamIsEq, AssertUnwindSafe, AtomicBool, AtomicPtr, Attr, auxtype, auxv, BackPlace, BacktraceContext, Barrier, BarrierWaitResult, Bencher, BenchMode, BenchSamples, BinaryHeap, BinaryHeapPlace, blkcnt, blkcnt64, blksize, BOOL, boolean, BOOLEAN, BoolTrie, BorrowError, BorrowMutError, Bound, Box, bpf, BTreeMap, BTreeSet, Bucket, BucketState, Buf, BufReader, BufWriter, Builder, BuildHasherDefault, BY, BYTE, Bytes, CannotReallocInPlace, cc, Cell, Chain, CHAR, CharIndices, CharPredicateSearcher, Chars, CharSearcher, CharsError, CharSliceSearcher, CharTryFromError, Child, ChildPipes, ChildStderr, ChildStdin, ChildStdio, ChildStdout, Chunks, ChunksMut, ciovec, clock, clockid, Cloned, cmsgcred, cmsghdr, CodePoint, Color, ColorConfig, Command, CommandEnv, Component, Components, CONDITION, condvar, Condvar, CONSOLE, CONTEXT, Count, Cow, cpu, CRITICAL, CStr, CString, CStringArray, Cursor, Cycle, CycleIter, daddr, DebugList, DebugMap, DebugSet, DebugStruct, DebugTuple, Decimal, Decoded, DecodeUtf16, DecodeUtf16Error, DecodeUtf8, DefaultEnvKey, DefaultHasher, dev, device, Difference, Digit32, DIR, DirBuilder, dircookie, dirent, dirent64, DirEntry, Discriminant, DISPATCHER, Display, Divdf3, Divdi3, Divmoddi4, Divmodsi4, Divsf3, Divsi3, Divti3, dl, Dl, Dlmalloc, Dns, DnsAnswer, DnsQuery, dqblk, Drain, DrainFilter, Dtor, Duration, DwarfReader, DWORD, DWORDLONG, DynamicLibrary, Edge, EHAction, EHContext, Elf32, Elf64, Empty, EmptyBucket, EncodeUtf16, EncodeWide, Entry, EntryPlace, Enumerate, Env, epoll, errno, Error, ErrorKind, EscapeDebug, EscapeDefault, EscapeUnicode, event, Event, eventrwflags, eventtype, ExactChunks, ExactChunksMut, EXCEPTION, Excess, ExchangeHeapSingleton, exit, exitcode, ExitStatus, Failure, fd, fdflags, fdsflags, fdstat, ff, fflags, File, FILE, FileAttr, filedelta, FileDesc, FilePermissions, filesize, filestat, FILETIME, filetype, FileType, Filter, FilterMap, Fixdfdi, Fixdfsi, Fixdfti, Fixsfdi, Fixsfsi, Fixsfti, Fixunsdfdi, Fixunsdfsi, Fixunsdfti, Fixunssfdi, Fixunssfsi, Fixunssfti, Flag, FlatMap, Floatdidf, FLOATING, Floatsidf, Floatsisf, Floattidf, Floattisf, Floatundidf, Floatunsidf, Floatunsisf, Floatuntidf, Floatuntisf, flock, ForceResult, FormatSpec, Formatted, Formatter, Fp, FpCategory, fpos, fpos64, fpreg, fpregset, FPUControlWord, Frame, FromBytesWithNulError, FromUtf16Error, FromUtf8Error, FrontPlace, fsblkcnt, fsfilcnt, fsflags, fsid, fstore, fsword, FullBucket, FullBucketMut, FullDecoded, Fuse, GapThenFull, GeneratorState, gid, glob, glob64, GlobalDlmalloc, greg, group, GROUP, Guard, GUID, Handle, HANDLE, Handler, HashMap, HashSet, Heap, HINSTANCE, HMODULE, hostent, HRESULT, id, idtype, if, ifaddrs, IMAGEHLP, Immut, in, in6, Incoming, Infallible, Initializer, ino, ino64, inode, input, InsertResult, Inspect, Instant, int16, int32, int64, int8, integer, IntermediateBox, Internal, Intersection, intmax, IntoInnerError, IntoIter, IntoStringError, intptr, InvalidSequence, iovec, ip, IpAddr, ipc, Ipv4Addr, ipv6, Ipv6Addr, Ipv6MulticastScope, Iter, IterMut, itimerspec, itimerval, jail, JoinHandle, JoinPathsError, KDHELP64, kevent, kevent64, key, Key, Keys, KV, l4, LARGE, lastlog, launchpad, Layout, Lazy, lconv, Leaf, LeafOrInternal, Lines, LinesAny, LineWriter, linger, linkcount, LinkedList, load, locale, LocalKey, LocalKeyState, Location, lock, LockResult, loff, LONG, lookup, lookupflags, LookupHost, LPBOOL, LPBY, LPBYTE, LPCSTR, LPCVOID, LPCWSTR, LPDWORD, LPFILETIME, LPHANDLE, LPOVERLAPPED, LPPROCESS, LPPROGRESS, LPSECURITY, LPSTARTUPINFO, LPSTR, LPVOID, LPWCH, LPWIN32, LPWSADATA, LPWSAPROTOCOL, LPWSTR, Lshrdi3, Lshrti3, lwpid, M128A, mach, major, Map, mcontext, Metadata, Metric, MetricMap, mflags, minor, mmsghdr, Moddi3, mode, Modsi3, Modti3, MonitorMsg, MOUNT, mprot, mq, mqd, msflags, msghdr, msginfo, msglen, msgqnum, msqid, Muldf3, Mulodi4, Mulosi4, Muloti4, Mulsf3, Multi3, Mut, Mutex, MutexGuard, MyCollection, n16, NamePadding, NativeLibBoilerplate, nfds, nl, nlink, NodeRef, NoneError, NonNull, NonZero, nthreads, NulError, OccupiedEntry, off, off64, oflags, Once, OnceState, OpenOptions, Option, Options, OptRes, Ordering, OsStr, OsString, Output, OVERLAPPED, Owned, Packet, PanicInfo, Param, ParseBoolError, ParseCharError, ParseError, ParseFloatError, ParseIntError, ParseResult, Part, passwd, Path, PathBuf, PCONDITION, PCONSOLE, Peekable, PeekMut, Permissions, PhantomData, pid, Pipes, PlaceBack, PlaceFront, PLARGE, PoisonError, pollfd, PopResult, port, Position, Powidf2, Powisf2, Prefix, PrefixComponent, PrintFormat, proc, Process, PROCESS, processentry, protoent, PSRWLOCK, pthread, ptr, ptrdiff, PVECTORED, Queue, radvisory, RandomState, Range, RangeFrom, RangeFull, RangeInclusive, RangeMut, RangeTo, RangeToInclusive, RawBucket, RawFd, RawHandle, RawPthread, RawSocket, RawTable, RawVec, Rc, ReadDir, Receiver, recv, RecvError, RecvTimeoutError, ReentrantMutex, ReentrantMutexGuard, Ref, RefCell, RefMut, REPARSE, Repeat, Result, Rev, Reverse, riflags, rights, rlim, rlim64, rlimit, rlimit64, roflags, Root, RSplit, RSplitMut, RSplitN, RSplitNMut, RUNTIME, rusage, RwLock, RWLock, RwLockReadGuard, RwLockWriteGuard, sa, SafeHash, Scan, sched, scope, sdflags, SearchResult, SearchStep, SECURITY, SeekFrom, segment, Select, SelectionResult, sem, sembuf, send, Sender, SendError, servent, sf, Shared, shmatt, shmid, ShortReader, ShouldPanic, Shutdown, siflags, sigaction, SigAction, sigevent, sighandler, siginfo, Sign, signal, signalfd, SignalToken, sigset, sigval, Sink, SipHasher, SipHasher13, SipHasher24, size, SIZE, Skip, SkipWhile, Slice, SmallBoolTrie, sockaddr, SOCKADDR, sockcred, Socket, SOCKET, SocketAddr, SocketAddrV4, SocketAddrV6, socklen, speed, Splice, Split, SplitMut, SplitN, SplitNMut, SplitPaths, SplitWhitespace, spwd, SRWLOCK, ssize, stack, STACKFRAME64, StartResult, STARTUPINFO, stat, Stat, stat64, statfs, statfs64, StaticKey, statvfs, StatVfs, statvfs64, Stderr, StderrLock, StderrTerminal, Stdin, StdinLock, Stdio, StdioPipes, Stdout, StdoutLock, StdoutTerminal, StepBy, String, StripPrefixError, StrSearcher, subclockflags, Subdf3, SubI128, SuboI128, SuboU128, subrwflags, subscription, Subsf3, SubU128, Summary, suseconds, SYMBOL, SYMBOLIC, SymmetricDifference, SyncSender, sysinfo, System, SystemTime, SystemTimeError, Take, TakeWhile, tcb, tcflag, TcpListener, TcpStream, TempDir, TermInfo, TerminfoTerminal, termios, termios2, TestDesc, TestDescAndFn, TestEvent, TestFn, TestName, TestOpts, TestResult, Thread, threadattr, threadentry, ThreadId, tid, time, time64, timespec, TimeSpec, timestamp, timeval, timeval32, timezone, tm, tms, ToLowercase, ToUppercase, TraitObject, TryFromIntError, TryFromSliceError, TryIter, TryLockError, TryLockResult, TryRecvError, TrySendError, TypeId, U64x2, ucontext, ucred, Udivdi3, Udivmoddi4, Udivmodsi4, Udivmodti4, Udivsi3, Udivti3, UdpSocket, uid, UINT, uint16, uint32, uint64, uint8, uintmax, uintptr, ulflags, ULONG, ULONGLONG, Umoddi3, Umodsi3, Umodti3, UnicodeVersion, Union, Unique, UnixDatagram, UnixListener, UnixStream, Unpacked, UnsafeCell, UNWIND, UpgradeResult, useconds, user, userdata, USHORT, Utf16Encoder, Utf8Error, Utf8Lossy, Utf8LossyChunk, Utf8LossyChunksIter, utimbuf, utmp, utmpx, utsname, uuid, VacantEntry, Values, ValuesMut, VarError, Variables, Vars, VarsOs, Vec, VecDeque, vm, Void, WaitTimeoutResult, WaitToken, wchar, WCHAR, Weak, whence, WIN32, WinConsole, Windows, WindowsEnvKey, winsize, WORD, Wrapping, wrlen, WSADATA, WSAPROTOCOL, WSAPROTOCOLCHAIN, Wtf8, Wtf8Buf, Wtf8CodePoints, xsw, xucred, Zip, zx}
, morekeywords=[5]{assert!, assert_eq!, assert_ne!, cfg!, column!, compile_error!, concat!, concat_idents!, debug_assert!, debug_assert_eq!, debug_assert_ne!, env!, eprint!, eprintln!, file!, format!, format_args!, include!, include_bytes!, include_str!, line!, module_path!, option_env!, panic!, print!, println!, select!, stringify!, thread_local!, try!, unimplemented!, unreachable!, vec!, write!, writeln!}  
}%

\lstdefinestyle{colouredRust}%
{ basicstyle=\ttfamily%
, identifierstyle=%
, commentstyle=\color[gray]{0.4}%
, stringstyle=\color[rgb]{0, 0, 0.5}%
, keywordstyle=\bfseries
, keywordstyle=[2]\color[rgb]{0.75, 0, 0}
, keywordstyle=[3]\color[rgb]{0, 0.5, 0}
, keywordstyle=[4]\color[rgb]{0, 0.5, 0}
, keywordstyle=[5]\color[rgb]{0, 0, 0.75}
, columns=spaceflexible%
, keepspaces=true%
, showspaces=false%
, showtabs=false%
, showstringspaces=true%
}%

\lstdefinestyle{boxed}{
  style=colouredRust%
, numbers=left%
, firstnumber=auto%
, numberblanklines=true%
, frame=trbL%
, numberstyle=\tiny%
, frame=leftline%
, numbersep=7pt%
, framesep=5pt%
, framerule=10pt%
, xleftmargin=15pt%
, backgroundcolor=\color[gray]{0.97}%
, rulecolor=\color[gray]{0.90}%
}

\lstnewenvironment{lstrust}{}{}

\title{\sysname: A Large-Scale Dataset for Smart Contract Auditing}

\author{
    \IEEEauthorblockN{
    Shihao Xia\IEEEauthorrefmark{1},  
    Mengting He\IEEEauthorrefmark{1}, 
Linhai Song\IEEEauthorrefmark{1},
Yiying Zhang\IEEEauthorrefmark{2}
} 
   
   \IEEEauthorblockA{
    {\textit{\IEEEauthorrefmark{1}The Pennsylvania State University} 
    \textit{\IEEEauthorrefmark{2}University of California, San Diego}}
    }
  \IEEEauthorblockA{ Email: 
    {\{sxia, mvh6224\}@psu.edu,
    songlh@ist.psu.edu,
    yiying@ucsd.edu
    }
    }
}

\def\BibTeX{{\rm B\kern-.05em{\sc i\kern-.025em b}\kern-.08em
    T\kern-.1667em\lower.7ex\hbox{E}\kern-.125emX}}
\begin{document}

\lstdefinelanguage{Solidity}{
  basicstyle=\ttfamily,
  stepnumber=1,
  numbersep=5pt,
  breaklines=true,
  showstringspaces=false,
  frame=single,
  tabsize=2,
  keywords=[1]{anonymous, assembly, assert, balance, break, call, callcode, case, catch, class, constant, continue, constructor, contract, debugger, default, delegatecall, delete, do, else, emit, event, experimental, export, external, false, finally, for, function, gas, if, implements, import, in, indexed, instanceof, interface, internal, is, length, library, log0, log1, log2, log3, log4, memory, modifier, new, payable, pragma, private, protected, public, pure, push, require, return, returns, revert, selfdestruct, send, solidity, storage, struct, suicide, super, switch, then, this, throw, transfer, true, try, typeof, using, value, view, while, with, addmod, ecrecover, keccak256, mulmod, ripemd160, sha256, sha3}, 
	keywordstyle=[1]\color{blue}\bfseries,
	keywords=[2]{address, bool, byte, bytes, bytes1, bytes2, bytes3, bytes4, bytes5, bytes6, bytes7, bytes8, bytes9, bytes10, bytes11, bytes12, bytes13, bytes14, bytes15, bytes16, bytes17, bytes18, bytes19, bytes20, bytes21, bytes22, bytes23, bytes24, bytes25, bytes26, bytes27, bytes28, bytes29, bytes30, bytes31, bytes32, enum, int, int8, int16, int24, int32, int40, int48, int56, int64, int72, int80, int88, int96, int104, int112, int120, int128, int136, int144, int152, int160, int168, int176, int184, int192, int200, int208, int216, int224, int232, int240, int248, int256, mapping, string, uint, uint8, uint16, uint24, uint32, uint40, uint48, uint56, uint64, uint72, uint80, uint88, uint96, uint104, uint112, uint120, uint128, uint136, uint144, uint152, uint160, uint168, uint176, uint184, uint192, uint200, uint208, uint216, uint224, uint232, uint240, uint248, uint256, var, void, ether, finney, szabo, wei, days, hours, minutes, seconds, weeks, years},	
	keywordstyle=[2]\color{teal}\bfseries,
	keywords=[3]{block, blockhash, coinbase, difficulty, gaslimit, number, timestamp, msg, data, gas, sender, sig, value, now, tx, gasprice, origin},	
	keywordstyle=[3]\color{violet}\bfseries,
	identifierstyle=\color{black},
	sensitive=true,
	comment=[l]{//},
	morecomment=[s]{/*}{*/},
	commentstyle=\color{gray}\ttfamily,
	stringstyle=\color{red}\ttfamily,
	morestring=[b]',
	morestring=[b]",
}

\lstdefinelanguage{GPTPrompt}{
basicstyle=\small,
  numbersep=5pt,
  breaklines=true,
  showstringspaces=false,
  breakindent=1em,
  frame=single,
  identifierstyle=\color{black}
}

\lstdefinelanguage{YAML}{
basicstyle=\small,
  numbersep=5pt,
  breaklines=true,
  showstringspaces=false,
  breakindent=1em,
  frame=single,
    keywords={true,false,null,y,n},
    keywordstyle=\color{darkgray}\bfseries,
    basicstyle=\ttfamily\footnotesize,
    sensitive=false,
    comment=[l]{\#},
    morecomment=[s]{/*}{*/},
    commentstyle=\color{purple}\ttfamily,
    stringstyle=\color{red}\ttfamily,
    morestring=[b]',
    morestring=[b]"
}

\maketitle

\begin{abstract}
  There is a huge demand to ensure the compliance of smart contracts listed on blockchain 
platforms to safety and economic standards described in natural languages. 
Today, manual efforts in the form of auditing are commonly used to achieve this goal. 
ML-based automated techniques have the promise to alleviate human efforts 
and the resulting monetary costs. However, unlike other domains where ML techniques have had huge successes, no systematic ML techniques have been 
proposed or applied to smart contract auditing. 
We present \sysname, the first dataset for automated smart-contract 
auditing research. \sysname\ consists of 5,377 real-world smart contracts 
running on Ethereum, a widely used blockchain platform, 
and 15,975 violations of standards on Ehereum called ERCs. Out of these violations, 139 are real violations programmers made. 
The remaining are errors systematically injected by us to reflect the violations of different ERC rules. 
We evaluate \sysname\ using GPT-4 by prompting it with both the contracts and ERC rules.
In addition, we manually identify each violated rule and the corresponding code site (\ie, oracle) and prompt GPT-4 with the information asking for a True-or-False question. Our results show that without the oracle, GPT-4 can only detect 0.9\% violations, 
and with the oracle, it detects 22.9\% violations. 
These results show the potential room for improvement in ML-based techniques for smart-contract auditing.

\end{abstract}


\section{Introduction}
\label{sec:intro}

Ethereum~\cite{eth-1,eth-2} is a decentralized, open-source blockchain platform that has become the de facto for running decentralized applications like smart contracts~\cite{dapps, sc-anatomy}. 
To standardize smart contract implementation,
Ethereum Request for Comments (ERCs) have been developed. Each ERC provides 
a set of formal standards and is typically written 
in natural languages~\cite{token-standard}.
For example, ERC20~\cite{erc20} defines the rules for fungible tokes  
--- digital assets that are interchangeable. 
ERCs are essential in the Ethereum ecosystem, 
offering a common framework that developers must follow when implementing smart contracts.
Violations of ERC rules can result in interoperability issues, 
contract failures, and financial loss. Moreover, tokens that don't comply 
with ERCs may be delisted from exchanges, as many exchanges require ERC compliance for listing~\cite{erc-standard}.

Despite the importance of adhering to ERC standards, 
developers often find it challenging to comply fully 
due to the complexity of the rules and their contract code. 
ERC standards consist of numerous rules, and this number continues 
to grow as new standards are introduced. For the three ERCs discussed in this work, 
there are 132 rules. These rules are presented in various formats, 
with some outlined as code comments 
and others explained in natural language paragraphs. 
Meanwhile, smart contract code is typically intricate, 
often spanning thousands of lines across multiple files. 
Some details may be hidden within complex caller-callee relationships, 
while others might involve objects and functions coded by different developers. 
This combination of ERC complexity and the intricacies of smart 
contracts makes it exceedingly difficult for programmers to ensure full 
compliance with ERC rules. 
Consequently, ERC violations are common in real-world smart contracts~\cite{humanaudited}.

To detect ERC violations, today's common practice heavily relies on human efforts.
Automated, program-analysis-based checkers for ERC rules do exist~\cite{slither-erc,erc20-verifier}, but they fail to detect complex violations because of many ERC rules' non-structured, natural-language-based definitions. 
As a result, smart contracts commonly undergo human auditing, provided by specialized services with security experts~\cite{certik,revoluzion,pixelplex,blockhunters, immunebytes, antier,humanaudited}. 
Auditing services are not only costly but also slow. 
For example, we examined the history of 30 smart contracts that were submitted for manual auditing on a platform called Ethereum Commonwealth Security Department~\cite{humanaudited}. 
We found that each contract only has an average of 260 lines of code but 
is audited for ten days with an estimated cost of \$500.
Clearly, manual auditing is not a scalable approach.

A promising, scalable approach for automated smart-contract auditing is to leverage large 
language models (LLMs), both because of LLMs' success in domains 
like program generation~\cite{singh2023progprompt} 
and bug fixing~\cite{jin2023inferfix, xia2023keep} and because of 
a fair amount of ERC rules' natural-language descriptions. To develop LLM-based techniques (\eg, fine-tuning, few-shot learning) 
for smart contract auditing, an important step is to construct quality datasets. 
Unfortunately, no smart contract datasets exist for ERC rule checking and fixing. Traditional program bug datasets~\cite{saha2018bugs, fan2020ac} cannot be used for smart-contract auditing, as unlike ERC rules, compiler errors and program runtime errors have well-defined, structured definitions.

To drive research and practices in smart-contract auditing and to assist real users with their auditing tasks, we release {\em \sysname}, the first dataset of real-world smart contracts and their ERC-rule violations. \sysname\ consists of \textbf{15,975 ERC violations and 5,377 real-world smart contracts}, collected from etherscan.io~\cite{etherscan}, polygonscan.com~\cite{polygonscan}, and Ethereum Commonwealth Security Department~\cite{humanaudited}. 139 violations from 30 contracts are real-world ERC violations we collect and inspect. 
As real violations are rarely published, we build program analysis techniques 
and inject 15,836 violations into 5,347 contracts according to 88 ERC rules. 

We evaluate \sysname\ using GPT-4 with two different approaches. 
First, we prompt GPT-4 with the contract to be inspected along with the full rule set of the corresponding ERC. 
Our results show that GPT-4 detects only 29\% of real violations 
and 0.6\% of injected violations. To assess how GPT-4 might perform with 
additional oracle information, we manually identify the violated ERC 
rules and the specific code sites causing the violations. 
We then prompt GPT-4 with this precise information and ask it a True-or-False question. 
In this case, GPT-4 successfully detects 42.8\% of real violations 
and 22.8\% of injected violations. The improved detection rates highlight the potential room 
for machine learning-based techniques in automating smart contract audits.

Overall, \sysname\ goes beyond being a key contribution to Ethereum smart contract auditing. Other software auditing and verification tasks, 
such as ensuring the compliance of API usage rules, 
can potentially use \sysname's contract dataset, its violation dataset, 
or its methodologies for evaluation.

We have released our dataset, results, and source code of our injecting scripts, all of which
can be found at \url{https://github.com/charlesxsh/scbench}.


\section{Background}
This section provides background on Ethereum, smart contracts, ERCs, and ERC violations.

\subsection{Ethereum and Smart Contracts}

Ethereum is a blockchain platform where developers can create and deploy smart contracts to build decentralized applications (dApps)~\cite{eth-1,eth-2}. 
Both Ethereum users and smart contracts have their own unique Ethereum addresses, 
which allow them to send and receive Ether (the native cryptocurrency of Ethereum) 
and interact with smart contracts to carry out complex transactions 
for a variety of purposes. Ethereum has grown into a thriving digital economy ecosystem, 
with a total market capitalization exceeding \$200 billion at the time of writing and more than one million transactions processed daily, 
amounting to over \$4 billion in volume~\cite{eth-price, eth-daily}. 
Smart contracts are central to Ethereum's success, driving the majority 
of transactions and powering key functionalities such as 
cryptocurrencies, NFTs, and decentralized finance (DeFi)~\cite{erc20,erc721,eth-defi}.

Smart contracts are commonly written in the Solidity programming language~\cite{solidity-popular-1,solidity-popular-2}. 
An example of a smart contract is presented 
in Figure~\ref{fig:20-high}. The contract has two contract fields in lines 2 and 3, 
\texttt{\_balances} (line 2) and \texttt{\_allowances} (line 3), 
tracking the number of tokens owned by each address and 
the tokens approved by the first dimension for manipulation 
by the second dimension, respectively. 
The function \texttt{transferFrom()} in lines 6--10 facilitates the transfer of 
\texttt{amount} tokens from one address to another. 
\texttt{transferFrom()} can be called by any Ethereum user or contract after the contract is deployed, while the internal function \texttt{\_transfer()} (lines 11--17) is restricted to calls from the same address.

{
\begin{figure}[t]

\begin{minipage}{\columnwidth}
\begin{center}
\scriptsize
\begin{lstlisting}[numbers=left,framexleftmargin=.15in,xleftmargin=.15in,language=Solidity,basicstyle=\ttfamily,morekeywords={-},morekeywords={+},keepspaces=true]
 contract ERC20 {
   mapping(address => uint256) _balances;
   mapping(address => mapping(address => uint256)) _allowances;
   event Transfer(address indexed _from, address indexed _to, uint256 _value);
   
   function transferFrom(address from, address to, uint256 amount) public returns (bool) {
+    _allowances[from][msg.sender] -= amount;
     _transfer(from, to, amount);
     return true;
   }
   function _transfer(address from, address to, uint256 amount) internal {
     require(from != address(0), "transfer from address zero");
     require(to != address(0), "transfer to address zero");
     _balances[from] -= amount;
     _balances[to] += amount;
     emit Transfer(from, to, amount);
   }
 }
\end{lstlisting}
\caption{An ERC20 rule violation that can be exploited to steal tokens.
{\textit{(The code is simplified for illustration purpose.)}}}
\label{fig:20-high}
\end{center}
\end{minipage}
\end{figure}
}

\subsection{Ethereum Request for Comment (ERC)}

ERCs are technical specifications that define the requirements for implementing smart contracts. Those requirements aim to ensure
compatibility across different contracts, applications, and platforms. 
By standardizing the contract implementations, 
ERCs help strengthen and promote the growth of the Ethereum ecosystem~\cite{erc-eip1, erc-standard, stefanovic2023proposal}. 

Typically, an ERC begins with a brief explanation of its motivation. For instance, ERC20~\cite{erc20} aims to establish 
a standard token interface that can be used by applications 
such as wallets and decentralized exchanges.
After the motivation, an ERC specifies all the necessary public functions 
and events by outlining their parameters, return values, and any optional attributes for the parameters. 
It also provides implementation requirements in the form of plain text or 
code comments for each function or event declaration. 
For example, besides the requirements for the function API and return value generation, ERC20 includes the following rules for the \texttt{transferFrom()} 
function (as shown in Figure~\ref{fig:20-high}), 
which mandate emitting a \texttt{Transfer} event, 
verifying that the message sender has been approved to manage 
the token owner’s tokens (and throwing an exception if not), 
treating the transfer of zero tokens in the same way as any other amount, 
and requiring an event to be emitted even when transferring zero tokens.

\subsection{ERC Rule Violations}

An ERC rule violation occurs when a smart contract is expected to follow a specific rule, but certain aspects of the contract do not. 
Figure~\ref{fig:20-high} illustrates an instance of an ERC20 rule violation in a real smart contract, where the \texttt{transferFrom()} function fails to check 
whether the caller has the necessary authorization to transfer 
the specified \texttt{amount} of tokens. 
This verification is required by ERC20 to ensure financial security. 
As a result of this oversight, anyone could potentially steal tokens 
from any address by invoking \texttt{transferFrom()} to transfer tokens to their own address.
The patch shown in line 7 offers a fix for this violation. 
It employs a two-dimensional map, \texttt{\_allowances}, 
to track the number of tokens the ``from’’ address allows 
\texttt{msg.sender} to manage. 
If the subtraction operation in this line results in an underflow, 
an exception is triggered, causing the transaction to terminate. 
This fix ensures that the message caller cannot transfer tokens 
unless they have enough privilege.

Violating ERC rules can lead to significant financial losses and unpredictable contract behavior. 
For example, ERC721 requires the \texttt{onERC721Received()} function to be called for each token transfer when the recipient is a contract. 
Additionally, it mandates that the caller must check 
if the return value of \texttt{onERC721Received()} matches a specific magic number. 
These two rules ensure that the recipient contract is capable of 
properly handling the transferred tokens.
If tokens are sent to a contract that cannot handle them, 
they can become permanently locked within the contract. 
This issue was first reported in 2017 on Ethereum Reddit, 
resulting in the loss of \$10,000 worth of tokens at the time, 
and it has since led to millions of dollars in losses~\cite{erc20-problem-history}. 
In short, ensuring contracts comply with ERC rules is essential to protect 
financial assets and maintain proper contract functionality.

\begin{table}[t]
\centering
\scriptsize
\vspace{2mm}
\caption{How ERC rules are distributed
across different error-injection methods. 
\textit{(``Uncovered'': rules whose violations cannot be injected
by our designed error-injection methods.)}}
\label{tab:rules}
\vspace{0.1in}
\setlength{\tabcolsep}{1.3mm}
\begin{tabular}{|l|c|c|c|c|c|c|c||c|}

\hline
 \textbf{ID} & \textbf{Check} & \textbf{API} & \textbf{Value} & \textbf{Call} & \textbf{Return} & \textbf{Logging} & \textbf{Uncovered}  & \textbf{Total}  \\ \hline \hline
ERC20 & 1 & 9 & 1 & 0 & 9 & 5 & 7 & 32         \\ \hline
ERC721 & 12 & 10 & 0 & 2 & 4 & 10 & 22 & 60      \\ \hline
ERC1155 & 7 & 6 & 0 & 2 & 0 & 7 & 18 & 40      \\ \hline \hline
Sum    & 20 & 25 & 1 & 4 & 13 & 22 & 47 & 132  \\
\hline
\end{tabular}
\end{table}

\subsection{Today's Auditing Practices}
The common practice for detecting ERC rule violations 
today relies on manual auditing, 
often provided 
by paid 
services~\cite{certik,revoluzion,pixelplex,blockhunters,immunebytes,antier,humanaudited}. One such service is the Ethereum Commonwealth Security Department~\cite{humanaudited}, where users submit smart contracts for auditing by filing a GitHub issue. 
The service then manually audits the submitted contracts 
and provides feedback through the issue.

To reduce the manual workload and associated costs, 
some automated tools have been developed using static program analysis. 
For example, Slither offers specific checkers (\ie, slither-check-erc~\cite{slither-erc}) 
that verify whether a given contract complies with the corresponding ERC standards 
for 11 ERCs. However, these tools have limited functionality. 
They primarily focus on ensuring the presence of required functions and events, 
confirming that these elements are correctly declared, 
and verifying that functions trigger the necessary events. 
Unfortunately, they are unable to check more advanced conditions, such as verifying whether the message caller has enough privilege to transfer tokens.
To our knowledge, no machine learning-based automated auditing 
techniques have been proposed yet. We hope that the release of \sysname\ 
can foster a new line of research in this regard.

\begin{table}[t]
\centering
\scriptsize
\vspace{2mm}
\caption{How many errors are injected by different error-injection methods.
}
\label{tab:inject}
\vspace{0.1in}
\setlength{\tabcolsep}{1.2mm}
\begin{tabular}{|l|c|c|c|c|c|c||c||c|}
\hline
 & \multicolumn{7}{c|}{\textbf{\# of Violations}} & \multirow{2}{*}{{\textbf{\# of Cont.}}} \\
 \cline{2-8}
& \textbf{Check} & \textbf{API} & \textbf{Value} & \textbf{Call} & \textbf{Return} & \textbf{Logging} & \textbf{Total}  &  \\ \hline \hline
ERC20        & 566 & 4605 & 736 & 0 & 5930 & 3612 & 15449 & 5211 \\ \hline
ERC721   & 158 & 72 & 0 & 15 & 48 & 33 & 326 & 110\\ \hline
ERC1155  & 18  & 30 & 0 & 4 & 9 & 0 & 61 & 26 \\ \hline \hline
Total    &  742   &  4707  & 736  &  19 & 5987 & 3645 & 15836 & 5347    \\ \hline

\end{tabular}
\end{table}

%
%

%
%
%

\section{\sysname}

This section outlines the process of building the dataset 
and provides some relevant statistics.

\subsection{Construction}

We collect real-world smart contracts from Ethereum Commonwealth Security Department~\cite{humanaudited}, etherscan.io~\cite{etherscan}, and polygonscan.com~\cite{polygonscan}.
As the purpose of \sysname\ is for evaluating automated ERC-rule checking, we include ERC rule violations of these collected contracts in \sysname\ using two approaches.
First, we manually inspect a set of real smart contracts and identify all their ERC violations.
This process is time-consuming, resulting in a limited number of violations. 
To address this, we perform automated error injection into a large number of real-world smart contracts, significantly increasing the number of violations. 
These two types of violations serve to validate each other 
and help ensure that the evaluated techniques demonstrate consistent performance.

\boldparagraph{Manual Inspection.}
We manually analyze 30 ERC20 contracts obtained from the Ethereum Commonwealth Security Department~\cite{humanaudited}, 
selecting the most recent 30 audit requests that meet the following criteria: 1) they contain Solidity source code, 
2) they have been approved by Solidity developers (as indicated by the ``approved’’ tag), 
3) they exhibit ERC rule violations, 
and 4) all contracts and the related Solidity code are within the same contract file. 
On average, each contract file contains 260.9 lines of Solidity code.
Through detailed examination, we identify 
139 ERC rule violations. 
 Of these, 27 violations present a clear method for exploitation that can lead 
to financial losses, and we classify them as having a high-security impact. 
Another 48 violations result from incorrect implementation of 
required functionalities, but there is no apparent way to exploit these 
for financial gain. Therefore, we consider them to have a medium-security impact. 
The remaining 64 violations involve failures in generating necessary logs 
and are categorized as having a low-security impact.

{
\begin{figure}[t]

\begin{minipage}{\columnwidth}
\begin{center}
\scriptsize
\begin{lstlisting}[numbers=left,framexleftmargin=.15in,xleftmargin=.15in,language=Solidity,basicstyle=\ttfamily,morekeywords={+},keepspaces=true]
 contract ERC1155 {
   function safeBatchTransferFrom(address from, address to, uint256[] memory ids, uint256[] memory amounts, bytes memory data) public  {
     require(from==_msgSender()||isApprovedForAll(from, _msgSender()), "not token owner or approved");
     _safeBatchTransferFrom(from,to,ids,amounts,data);
   }
   
   function _safeBatchTransferFrom(address from, address to, uint256[] memory ids, uint256[] memory amounts, bytes memory data) internal {
     require(ids.length==amounts.length, "ids and amounts length mismatch");
-    require(to!=address(0), "transfer to address 0");
     address operator = _msgSender();
     for (uint256 i = 0; i < ids.length; ++i) {
       uint256 id = ids[I];
       uint256 amount = amounts[I];
       _balances[id][from] -= amount;
       _balances[id][to] += amount;
     }
     emit TransferBatch(operator,from,to,ids,amounts);
  }
}
\end{lstlisting}
\caption{Violation injection of a condition-check rule.
{\textit{(Line 9 is deleted to perform the violation injection.)}}}
\label{fig:inj-exp}
\end{center}
\end{minipage}
\end{figure}
}

\boldparagraph{Error Injection.}
To augment the limited manual-inspected real violations, we perform error injection to 5,347 real-world smart contract source code using the following methodology. 
We design six error injection methods corresponding to 85 ERC rules, resulting in a total of 15,836 ERC violations. 
Table~\ref{tab:rules} shows the number of rules each method covers.

We focus on contracts in three ERC standards, ERC20~\cite{erc20}, ERC721~\cite{erc721},
and ERC1155~\cite{erc1155}, for two reasons.
First, these ERCs are significant and have numerous crucial financial applications. For instance, there are over 450,000 ERC20 tokens on the Ethereum platform~\cite{erc20-popular}, many of which have market capitalizations exceeding \$1 billion (\eg, USDT~\cite{USDT}, SHIB~\cite{SHIB}, Binance USD~\cite{Binance}).
Second, these three standards are among the most mature ERCs and inspire subsequent ERCs, making their rules representative of rules in other ERCs. 
As shown in Table~\ref{tab:rules}, these three ERCs specify 132 rules. Our error injection methods can cover 85 of them. We do not include the rest either because they are not clearly specified (\eg, ``throw if any other error'' in ERC1155) or because they require more complex static analysis or injection methods.

We collect contracts in these three ERCs from 
etherscan.io~\cite{etherscan} and polygonscan.com~\cite{polygonscan}.
These two platforms are the most popular analytics 
platforms for Ethereum and its sidechain Polygon~\cite{polygon}, respectively.
As shown in Table~\ref{tab:inject}, we collected 5,211 contracts that are supposed to target ERC20, 110 for ERC721, and 26 for ERC1155\footnote{We mainly rely on a contract’s name or the name of a base contract inherited by the contract 
to determine whether the contract implements a particular ERC.
}. 
On average, each of these contracts contains 477.5 lines of code.
We collect more contracts targeting ERC20 than ERC721 and ERC1155 contracts due to the significantly higher prevalence of ERC20 in practice.

We randomly inject one to three errors in each contract based on the fit of rules to the contract to construct ERC violations. 
To inject an error, we first convert each input smart contract source-code file into its abstract syntax tree (AST). 
We then randomly select a rule and apply the corresponding error-injection method 
to the AST. 
After modifying the AST, we convert it back into source code. 
We perform error injections on ASTs because it is easier to conduct static program analysis and modify code on ASTs than to work directly with the source code.
Table~\ref{tab:inject} shows the number of errors injected by each method. 
Finally, we verify that the modified source code compiles without errors using the Solidity compiler. 
Below, we discuss each error injection method in detail.

{
\begin{figure}[t]

\begin{minipage}{\columnwidth}
\begin{center}
\scriptsize
\begin{lstlisting}[numbers=left,framexleftmargin=.15in,xleftmargin=.15in,language=Solidity,basicstyle=\ttfamily,morekeywords={-},morekeywords={+},keepspaces=true]
contract ERC20 {
  mapping(address => uint256) _balances;
  
  function balanceOf(address account) public view returns (uint256) {
-   return _balances[account];
+   return _balances[account]+827;    
  }
}
\end{lstlisting}
\caption{Violation injection of a return rule.
{\textit{(Line 5 is replaced with line 6 to perform the violation injection.)}}}
\label{fig:inj-exp-erc20}
\end{center}
\end{minipage}
\end{figure}
}

\italicparagraph{Violations of Condition-Check Rules.}
ERCs mandate certain \texttt{public} functions to perform condition checks 
on their input parameters or callers’ addresses (\ie, \texttt{msg.sender}). 
Sometimes, these checks validate the input parameters. 
For example, ERC721 disallows the input parameter of function \texttt{ownerOf()} to be zero. 
More crucially, these checks ensure the caller has the permission to perform an action. 
For instance, ERC20’s \texttt{transfer(address \_to, uint256 \_value)} 
function sends tokens in the amount 
specified by its second parameter from the caller’s account to the receiver, 
whose address is the first parameter. 
ERC20 requires verifying that the caller has enough tokens 
(an amount greater than or equal to the second parameter) and throwing 
an exception if this condition is not met.

Many of these rules are enforced using \texttt{require} statements. 
To inject an error of this type into a function, we remove all \texttt{require} statements 
that check the parameter required by the rule as part of their conditions. 
Figure~\ref{fig:inj-exp} illustrates an example of how a rule violation is injected in this way.
ERC1155 requires that function \texttt{safeBatchTransferFrom()} ``MUST revert if `\_to’ 
is the zero address,'' where \texttt{\_to} is the second parameter of the function. 
Violating this rule would result in a permanent loss of the transferred tokens.
The violation injection method begins with the \texttt{safeBatchTransferFrom()} 
function in line 2, as the rule applies to this function. It ignores the \texttt{require} 
statement in line 3, since its conditions do not involve the \texttt{\_to} parameter. 
The method then proceeds to analyze the callee function \texttt{\_safeBatchTransferFrom()} defined in lines 7--18.
Similar to line 3, the method retains the \texttt{require} statement in line 7. 
However, it removes the \texttt{require} statement in line 9, 
as this one checks the \texttt{\_to} parameter. 
None of the remaining lines contain \texttt{require} statements, so they are left unchanged.
As shown in Figure~\ref{fig:inj-exp}, the modified contract can still be 
compiled by the Solidity compiler, but it allows \texttt{safeBatchTransferFrom()} to transfer tokens to the zero address, thus violating the rule.


\begin{figure*}[t]
   \begin{minipage}[t]{0.27\textwidth}
     \centering
     \includegraphics[width=\linewidth]{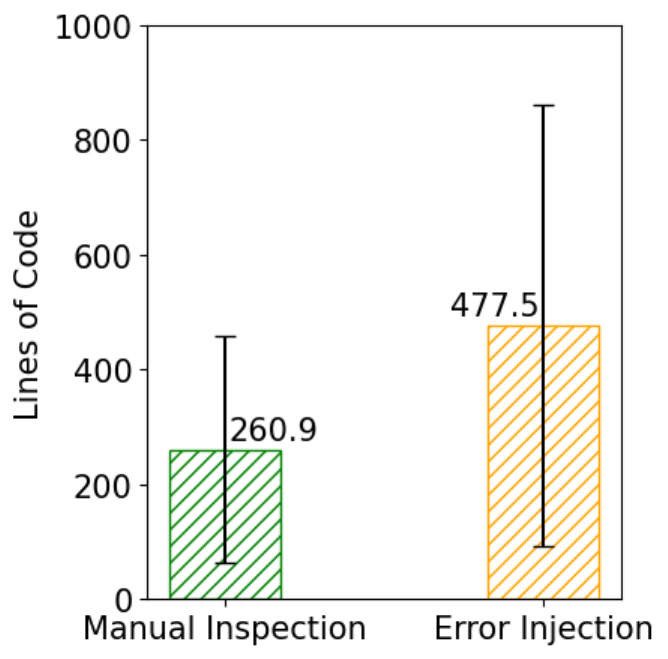}
     \caption{Average contract size.}\label{fig:size}
   \end{minipage}
   \hfill
   \begin{minipage}[t]{0.245\textwidth}
     \centering
     \includegraphics[width=\linewidth]{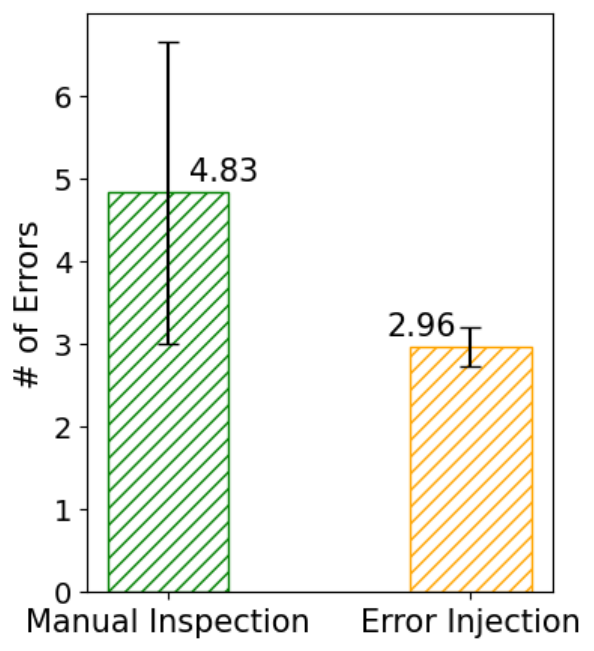}
     \caption{Average error number in a contract.}\label{fig:enum}
   \end{minipage}
   \hfill
   \begin{minipage}[t]{0.32\textwidth}
     \centering
     \includegraphics[width=\linewidth]{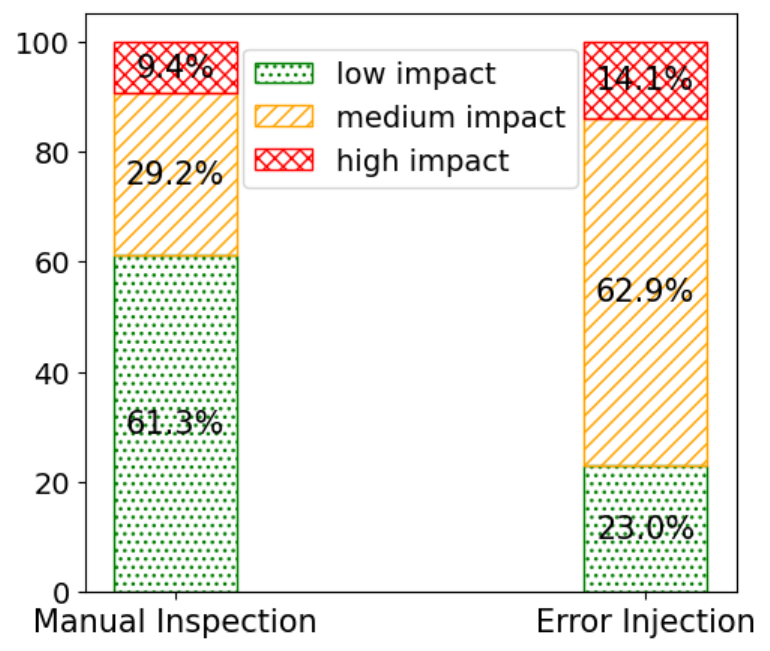}
     \caption{How errors distribute across different security impacts.}
\label{fig:impact-dis}
   \end{minipage}
\end{figure*}

\italicparagraph{Violations of API Rules.}
Each ERC specifies a set of required function call APIs that all contracts following the ERC must include. 
To inject an error for a required API, we identify and remove the function's definition, including both its declaration and its function body. 
To ensure the contract can still be compiled, 
we also remove all call sites of the function. 

\italicparagraph{Violations of Return-Value Rules.}
ERCs explicitly define how return values should be computed for certain function call APIs. For example, ERC20 requires that 
\texttt{allowance(address \_owner, address \_spender)} 
returns the token amount that \texttt{\_owner} 
allows \texttt{\_spender} to withdraw. 
As return-value rules apply to only four data types, integer, Boolean, address, 
and string,
we design four distinct error-injection methods for
each of them. We add a random integer to an integer return value, 
flip a Boolean return value, replace an address return with a random address, 
and change a string return 
value to an empty string if it is not already empty or replace an empty return
string with a random string.

Figure~\ref{fig:inj-exp-erc20} illustrates an example of injecting a violation of a rule that specifies how to generate a integer return value.
According to ERC20, the function \texttt{balanceOf(address \_owner)} must return ``the account balance of the another account with address \_owner.’’ 
The contract in Figure~\ref{fig:inj-exp-erc20} adheres to this rule, as shown in line 5.
To inject a violation, we modify the return value by adding a random number to it, 
as demonstrated in line 6.

\italicparagraph{Violations of Value-Update Rules.}
Fields or state variables in a contract represent its state. 
Some ERC rules specify how a public function should update a state variable. 
For example. contracts following ERC20 rules use 
a two-dimensional \texttt{mapping} to record the number of tokens the first key allows the second key to withdraw. 
The public function \texttt{approve(address \_spender, uint256 \_value)}  
sets the number 
of tokens (specified by the second parameter \texttt{\_value}) 
that the caller's 
address (\ie, the message sender) permits the first address parameter \texttt{\_spender} to withdraw. 
Therefore, ERC20 mandates that \texttt{approve(address \_spender, uint256 \_value)} updates the appropriate field 
of the two-dimensional \texttt{mapping}  with \texttt{\_value}.
To inject an error of this type into a function, 
we remove all assignments to the corresponding state variable within the function. 
The challenge is that ERCs do not prescribe how to 
name state variables, as these variables are only accessible within a contract. 
Consequently, given a rule of this type, different contracts are likely to use different names for the corresponding state variable.
Fortunately, ERCs require a getter function to return its value for most state variables, and this getter function has the same name across all contracts for each ERC. This requirement enables us to locate the correct state variable automatically and perform the error injection.

\italicparagraph{Violations of Function-Call Rules.}
Some ERC rules specify that a function must be called after a certain event. 
For instance, ERC721 requires \texttt{onERC721Received()} to be called if tokens are sent
to a contract. 
To inject an error for a call rule about function \texttt{A()} inside function \texttt{B()},
we simply remove all call sites of \texttt{A()} in \texttt{B()}.

\italicparagraph{Violations of Logging Rules.}
ERCs require specific events to be emitted within certain 
functions or after particular code actions for logging purposes. To inject such 
an error that violates one of those rules within a function, we remove all 
code statements within the function that emit the corresponding event.

\subsection{Dataset Summary}

Our dataset contains 15,975 ERC rule violations. Among them, 139 
are introduced by the real-world programmers, 
while the remaining 15,836 are injected by us. 
39 errors made by real-world programmers cannot be replicated using the error-injection methods. 
These include 28 errors that violate the rule that transferring zero tokens must be treated the same as transferring non-zero tokens and 11 errors that violate the rule that the \texttt{transfer()} function must throw an exception if the sender does not have enough balance.

There are 5,377 contracts in our dataset, 
including 30 originally ERC-violating contracts and 5,347 contracts with injected errors. 
On average, each contract contains 476.29 lines of code and 2.97 errors. 
Among these contracts, 5,241 contracts implement ERC20, 110 contracts implement ERC721, 
and 26 contracts implement ERC1155. 
ERC20 has the most contracts since it is the most popular ERC. 


Figures~\ref{fig:size} and Figure~\ref{fig:enum} compare errors from two sources on their contract sizes and the average number of errors per contract.
On average, an originally violating contract contains 260.9 lines of 
source code, with a standard deviation of 198, 
whereas a contract with injected errors contains 477.5 lines 
of code, with a standard deviation of 385. 
Each originally violating contract contains 4.83 errors on average, and each modified contract contains 2.96 injected errors on average.

The security impact of the errors is tied to the methods used for error injection. 
A total of 3,645 errors arise from failures to emit events, 
resulting in contract activities going unlogged. 
These are considered to have a low-security impact. Among the 2,230 
high-security impact errors, a significant number are caused by missing 
required checks to verify sufficient privileges for performing specific actions. 
The lack of these checks can be easily exploited, leading to financial losses (\eg, 
stolen tokens, as illustrated in Figure~\ref{fig:20-high}). 
These errors are categorized as having a high-security impact. 
Additionally, failures to call required functions or incorrect updates to state variables 
can also have a high-security impact. 
The remaining 9,961 errors cause contracts to behave unpredictably 
for users, although they may not directly result in financial loss, and are classified as having a medium-security impact. All these errors are injected through API, Value, Call, 
and Return methods.

Figure~\ref{fig:impact-dis} shows the distribution of errors across different security impacts for the two sources. For errors identified through manual inspection, the proportions of high-impact, medium-impact, and low-impact errors are 9.4\%, 29.2\%, and 61.3\%, respectively. 
For injected errors, the proportions are 14.1\%, 62.9\%, and 23.0\%, respectively.

\section{Evaluation}

{
\begin{figure}[t]

\begin{minipage}{\columnwidth}
\begin{center}
\scriptsize
\begin{lstlisting}[numbers=left,framexleftmargin=.15in,xleftmargin=.15in,language=GPTPrompt,basicstyle=\ttfamily,morekeywords={-},morekeywords={+},keepspaces=true]
The following code is the implementation of <ERC_type>. The <ERC_type> rules are 
attached below the code. Does this implementation violate <ERC_type> rules? 
Return a JSON array containing JSON objects with 'rule' and 'function' as keys, 
indicating the specific rule content that is violated and the function where the 
violation resides. 
code:"""
<code>
"""
<ERC_type>:
<ERC_content>
\end{lstlisting}
\caption{The full-rule prompting template.
{\textit{(For a concrete prompt, <ERC\_type> could be ERC20, ERC721, and ERC1155,
<code> is replaced with the whole contract code,
and <ERC\_content> is replaced with the whole ERC document.)}}}
\label{fig:prompt}
\end{center}
\end{minipage}
\end{figure}
}

{
\begin{figure}[t]

\begin{minipage}{\columnwidth}
\begin{center}
\scriptsize
\begin{lstlisting}[numbers=left,framexleftmargin=.15in,xleftmargin=.15in,language=GPTPrompt,basicstyle=\ttfamily,morekeywords={-},morekeywords={+},keepspaces=true]
Whether the following smart contract <contract_name> violate ERC rule <rule> for function <function_sig>? Answer YES or NO.
Code:"""
<code>
"""
\end{lstlisting}
\caption{The oracle prompting template.
{\textit{(For a concrete prompt, <contract\_name>, <rule>, <function\_sig>,
and <code> are replaced with
the name of the smart contract, the rule's natural language description, the declaration of the function, and the code of the function and all its callees.)}}}
\label{fig:prompt2}
\end{center}
\end{minipage}
\end{figure}
}

{
\begin{figure}[t]

\begin{minipage}{\columnwidth}
\begin{center}
\scriptsize
\begin{lstlisting}[numbers=left,framexleftmargin=.15in,xleftmargin=.15in,language=GPTPrompt,basicstyle=\ttfamily,morekeywords={-},morekeywords={+},keepspaces=true]
[
    {
        "rule": "approve: Allows _spender to withdraw from your account multiple times, up to the _value amount. If this function is called again it overwrites the current allowance with _value.",
        "function": "approve function is missing in the contract"
    },
    {
        "rule": "Approval: MUST trigger on any successful call to approve(address _spender, uint256 _value).",
        "function": "Approval event is missing in the contract"
    }
]
\end{lstlisting}
\caption{The GPT response for the contract in Figure~\ref{fig:20-high} with full-rule prompting.}
{}
\label{fig:20-high-prompt1-res}
\end{center}
\end{minipage}
\end{figure}
}

\begin{table*}[t]
\centering

\scriptsize
\vspace{2mm}
\caption{Experimental results for full-rule prompting. \textit{((x, y, z): x cases where both the reported rule and the reported violating function are correct, y cases where only the reported rule is correct, z cases where neither the rule nor the function is correct. ``-'': all numbers are zeros.)}}
\label{tab:prompt1}
\vspace{0.1in}
{
\setlength{\tabcolsep}{1.5mm}{
\begin{tabular}{|l|c|c|c|c|c|c|c|c|c|}
\hline
\multirow{2}{*}{{\textbf{ERC}}} & \multicolumn{4}{c|}{\textbf{Manual Inspection}} & \multicolumn{4}{c|}{\textbf{Error Injection}} & \multirow{2}{*}{{\textbf{Total}}}
       \\ \cline{2-9}
     & \multicolumn{1}{c|}{{\textbf{High}}} & {{\textbf{Medium}}} & {{\textbf{Low}}} & {{\textbf{Total}}} & {{\textbf{High}}} & {{\textbf{Medium}}} & {{\textbf{Low}}} & {{\textbf{Total}}} &
  \\ \hline \hline
ERC20              & (16,0,40)  & (21,0,36)   & (5,2,6)  & (42,2,82)  &  (3,0,3659)  & (83,5,7381)    &  (16,0,533) &  (102,5,11573) & (144, 7, 11655)\\ \hline
ERC721             &  - &  -  & -  & - & (0,0,108) & (0,0,195)   & (0,0,17) & (0,0,320) & (0,0,320)\\ \hline
ERC1155            &  - &  -  & -  & - & (0,0,4) & (0,0,31)   & (0,0,8) & (0,0,43) & (0,0,43) \\ \hline \hline
Total              & (16,0,40)  & (21,0,36)   & (5,2,6)  & (42,2,82)  & (3,0,3771) & (83,5,7607) & (16,0,558) & (102,5,11936) & (144, 7, 12018) \\ \hline
\end{tabular}
}
}

\end{table*}

\begin{table*}[t]
\centering

\scriptsize
\vspace{2mm}
\caption{Experimental results for oracle prompting. \textit{((x, y): x true positives, and y false negatives. ``-'': all numbers are zeros.)}}
\label{tab:prompt2}
\vspace{0.1in}
{
\setlength{\tabcolsep}{1.5mm}{
\begin{tabular}{|l|c|c|c|c|c|c|c|c|c|}
\hline
\multirow{2}{*}{{\textbf{ERC}}} & \multicolumn{4}{c|}{\textbf{Manual Inspection}} & \multicolumn{4}{c|}{\textbf{Error Injection}} & \multirow{2}{*}{{\textbf{Total}}}
       \\ \cline{2-9}
     & \multicolumn{1}{c|}{{\textbf{High}}} & {{\textbf{Medium}}} & {{\textbf{Low}}} & {{\textbf{Total}}} & {{\textbf{High}}} & {{\textbf{Medium}}} & {{\textbf{Low}}} & {{\textbf{Total}}} &
  \\ \hline \hline
ERC20              & (3,24)   & (29,19)   & (30,34)  & (62,77) & (739,1273) & (2037,7788)   & (823,2789)  & (3599,11850) & (3661,11927)\\ \hline
ERC721             &  - &  -  & -  & - & (2, 185)  & (1,105)   & (0,33) & (3,323) & (3,323)\\ \hline
ERC1155            &  - &  -  & -  & - &  (4,27) &  (0,30)  & - & (4,57)  & (4,57)\\ \hline \hline
Total              & (3,24)  & (29,19)   & (30,34)  & (62,77)  & (745,1485) & (2038,7923)   & (823, 2822) &  (3606, 12230) & (3668,12307)\\ \hline
\end{tabular}
}
}

\end{table*}

This section presents our evaluation results of \sysname\ using GPT-4.
Our experiments are designed to answer the following research questions: 
1) Coverage: How many errors can GPT-4 detect? and 2)
Accuracy: How accurate are GPT-4’s detection results?

\subsection{Methodology}

We evaluate \sysname\ using GPT-4 via the OpenAI API access.
We set its temperature value to zero to ensure that GPT-4's results are deterministic, enabling others to replicate our findings. 
Below, we detail our evaluation methodology.

For each auditing request, we ask GPT whether a contract violates any ERC rules, and if so, which function causes the violation.
To instruct GPT on ERC rules, we adopt in-context learning by providing GPT ERC rules in addition to the contract under inspection. 
We use two methodologies to provide ERC rules.
The first presents an ERC's official document with all rules to GPT,
as shown in Figure~\ref{fig:prompt}.
After GPT returns its output, we compare both the GPT-generated violating rules and violating functions to the ground truth.
We report when both are correct, when only the violating rule is correct, and when neither is correct.

The second method assumes oracle knowledge of which specific ERC rule(s) a contract violates and the function where the violation happens. 
So, it directly provides both to GPT and asks GPT a simple True-or-False question, as shown by Figure~\ref{fig:prompt2}.
A True result is correct (True Positive), while a False is wrong (False Negative).
For this method, we manually select rules and violation code sites based on either the original contract violations or our injected errors.

Presenting whole ERC rules with the contract serves as a baseline, while hand-picked rules can be viewed as an oracle.
Note that we do not provide more advanced methodology such as Chain-of-Thoughts~\cite{yao2022react}, as the focus of this work is to present our collected dataset and demonstrate its potential use.
Future research can explore the room for accuracy improvements.

\subsection{Experimental Results}

\boldparagraph{Full-Rule Prompting.}
As shown in Table~\ref{tab:prompt1}, using full-rule prompting, 
GPT-4 successfully detects 144 errors (0.9\% of the 15,975 errors), 
providing both the correct 
violated rules and the violating functions. 
For another 7 errors (0.04\%), 
GPT-4 only reports the correct 
rule but fails to identify the correct location. 
For the remaining 15,824 errors (99\%), 
GPT-4 does not provide any correct information.

For instance, Figure~\ref{fig:20-high-prompt1-res} illustrates 
GPT-4’s response when analyzing the contract in Figure~\ref{fig:20-high} 
using full-rule prompting. GPT-4 accurately identifies two rule violations 
and the functions responsible for them. Since the contract does not 
include the function \texttt{approve(address \_spender, uint256 \_value)}, it is considered violating
both that the function should overwrite the current allowance with \texttt{\_value} 
and that the function should emit the \texttt{Approval} event. 
However, GPT-4 fails to recognize that the contract fails to 
ensure the \texttt{transferFrom()} function properly validates its caller’s privileges, 
missing a critical violation with a high security impact. 
In another example, when asked to audit the contract in Figure~\ref{fig:inj-exp-erc20}, 
GPT-4 fails to detect that the \texttt{balanceOf(address account)} function 
does not return the expected value as required by ERC20.

We further separate the results between errors identified through manual inspection 
and those that are injected. For manually inspected errors, 
GPT-4 correctly reports both the violated rule and 
the violating function in 30\% of the 139 errors. 
For injected errors, this proportion is 0.6\%. 
The difference is probably due to that the injected errors 
are more challenging to identify than those introduced by programmers.

For errors in different security impacts, 
we notice GPT-4 has a very high detection rate when pinpointing 
manually introduced errors with a high security impact, with a detection rate to be 59.2\%.
For errors in other security impacts, 
the detection rate ranges from 0.13\% to 43.8\%. 


We then examine whether GPT-4's ability to detect violations varies
across different ERCs. We found that the detection rate for errors violating an ERC20 rule (0.9\%) is higher than for ERC721 (0\%) and ERC1155 (0\%). 
Notably, GPT-4 does not identify any errors in contracts 
implementing ERC721 and ERC1155. This discrepancy is likely because there 
are significantly more ERC20 contracts, 
meaning GPT-4 has probably been trained on a larger dataset of ERC20 contracts.

\boldparagraph{Oracle Prompting.}
Table~\ref{tab:prompt2} shows the results with oracle prompting. 
Under this setting, GPT-4's performance significantly improves, 
with the detection rate increasing from 0.9\% to 22.9\%. 
The improvement varies depending on the source of the errors. 
For manually inspected errors, the detection rate rises from 30\% to 44.6\%, 
while for injected errors, it increases from 0.6\% to 22.8\%. 
This boost in detection is mainly due to breaking down a complex task into smaller, 
more manageable tasks, allowing GPT-4 to focus on each one individually.
Those results show that simple prompting techniques even with oracle still have a large room for accuracy improvements.

For example, when presenting both the source code of the 
\texttt{transferFrom(address from, address to, uint256 amount)} function 
and its callee from lines 6 to 17 in Figure~\ref{fig:20-high}, along with 
the rule description stating that the function ``should throw unless the \texttt{\_from}
account has deliberately authorized the sender of the message via some mechanism,” 
and asking GPT-4 whether the rule is violated, GPT-4 correctly answers yes. 
In another example, when showing the \texttt{balanceOf()} function 
from lines 4 to 7 in Figure~\ref{fig:inj-exp-erc20}, along with the rule 
that the function “returns the account balance of another account with 
address \_owner,’’ and asking GPT-4 to analyze if the rule is violated, 
it correctly responds yes.

\section{Discussion and Conclusion}

We present \sysname, the first dataset for smart contract auditing. \sysname\ contains 5,377 real-world smart contracts and 15,975 ERC violations, with two sources of violations: real-world errors and injected errors. Our evaluation of \sysname\ with GPT-4 shows that while ML-based techniques are promising, there is still huge room for improvement.

Notably, \sysname\ has certain limitations. For example, we only include the three most important ERC standards and only inject one to three errors per contract. Both adding more ERC standards and supporting more types of error injection are 
feasible --- something we leave for future work. Another limitation is the imbalance of real-world violations and injected errors. As real-world violations are rarely reported, we believe the imbalance cannot be easily improved. One possible approach is to instruct ML models to create real-world-like errors.

\bibliography{gpt-1}
\bibliographystyle{abbrv}

\end{document}